\documentclass[twocolumn,amsmath,amssymb,osajnl]{revtex4}
\usepackage{epsfig}
\begin{document}

\title{Experimental investigation of cut-off phenomena in non-linear photonic crystal fibers}

\author{Jacob Riis Folkenberg and Niels Asger Mortensen}
\affiliation{Crystal Fibre A/S, Blokken 84, DK-3460
Birker\o d, Denmark}

\author{Kim P. Hansen and Theis P. Hansen}
\affiliation{Crystal Fibre A/S, Blokken 84, DK-3460 Birker\o d, 
Denmark\\COM, Technical University of Denmark, DK-2800 Kongens Lyngby, Denmark}

\author{Harald R. Simonsen and Christian Jakobsen}
\affiliation{Crystal Fibre A/S, Blokken 84, DK-3460 Birker\o d, Denmark~}

\begin{abstract}
The modal cut-off is investigated experimentally in a series of high quality non-linear photonic crystal fibers. We demonstrate a suitable measurement technique to determine the cut-off wavelength and verify it by inspecting the near field of the modes that may be excited below and above the cut-off. We observe a double peak structure in the cut-off spectra, which is attributed to a splitting of the higher order modes. The cut-off is measured for seven different fiber geometries with different pitches and relative hole size, and a very good agreement with recent theoretical work is found.
\end{abstract}


\maketitle

Since the invention of the photonic crystal fibers\cite{knight1996} several new fiber applications have emerged (for a recent review we refer to Ref.~\onlinecite{russell2003} and references therein). Especially, the use of small-core non-linear photonic crystal fibers (NL-PCFs) for super continuum generation\cite{ranka2000} and other non-linear processes has attracted much attention, both scientifically but also from an application point of view. Most of the pioneering experiments have been carried out on fibers with a small silica core surrounded by one or a few periods of large holes, and issues such as the regularity of the hole structure and the number of guided modes have not been considered in detail. However, with recent improvements in the production of NL-PCFs very large and regular holey structures are routinely achieved, and a comparison between theory and experiment is facilitated. One of the most important properties of fundamental nature is the modal cut-off of NL-PCFs, and its dependence on the pitch and hole size of the structure. In the present Letter we report on what we believe to be the first systematic experimental investigation of the modal cut-off in a series of NL-PCFs, and discuss the results in relation to recent theoretical work.\cite{mortensen2002a,kuhlmey2002,mortensen2003c}

\begin{figure}[b!]
\begin{center}
\epsfig{file=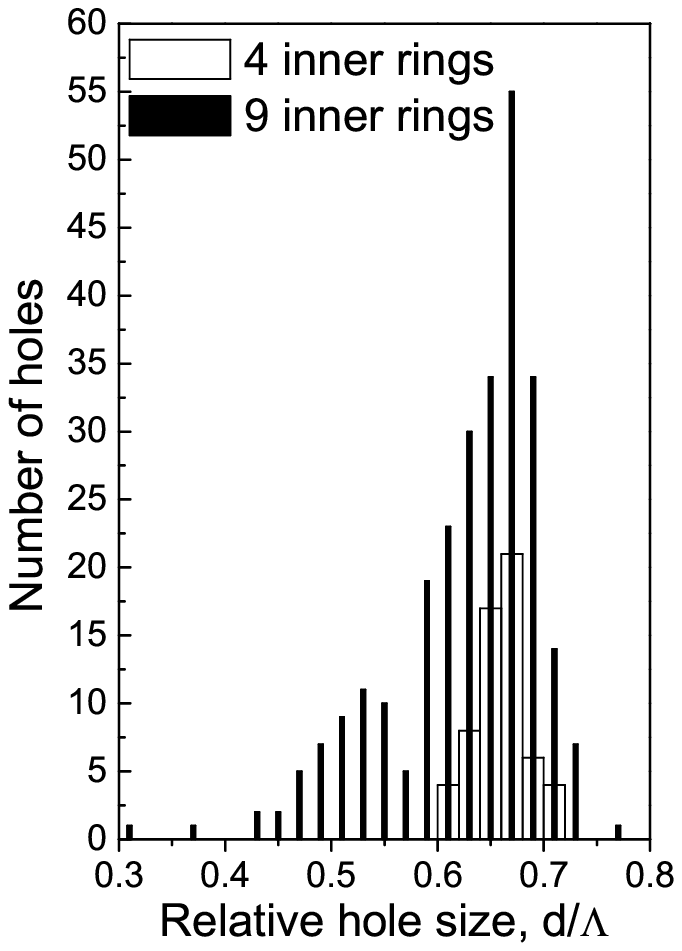, width=0.24\textwidth,clip}\epsfig{file=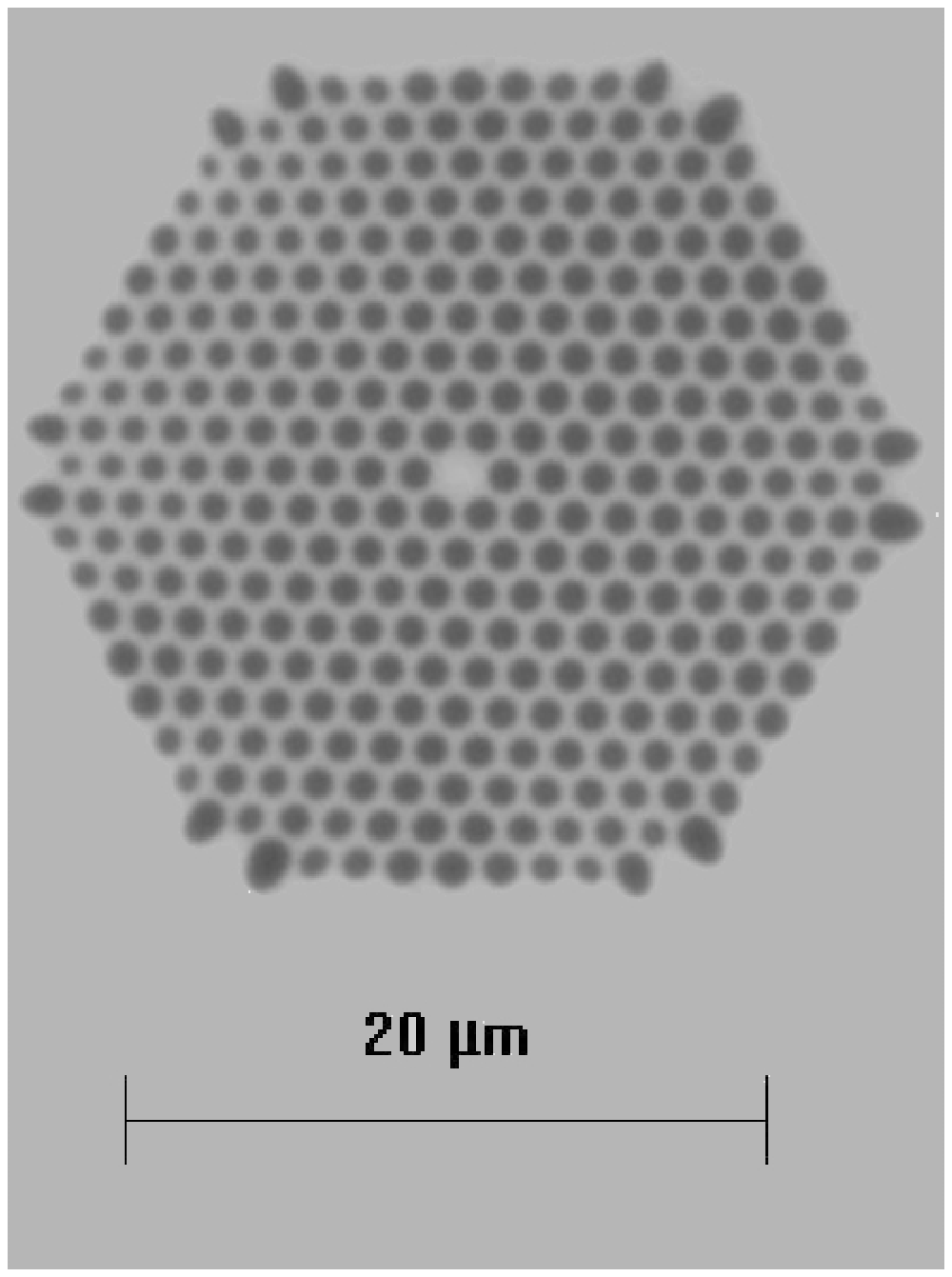, width=0.24\textwidth ,clip}
\end{center}
\caption{(Left) Histogram of the relative hole size distribution in the inner four and nine rings of the cladding. (Right) Scaled microscope picture of fiber \#1, see Table~\ref{tab1} for the spatial dimensions.}
\label{fig1}
\end{figure}

The fibers under investigation are fabricated using the typical PCF design with a triangular arrangement of air holes in the cladding and one missing hole that forms the core. All fibers have been drawn from the same preform having nine full rings of air holes and a tenth incomplete ring, see Fig.~\ref{fig1} (right). The air-hole diameter $d$ and air-hole pitch $\Lambda$ have been determined using a calibrated high-resolution optical microscope and commercially available image analysis software. Also shown in Fig.~\ref{fig1} (left) is a histogram of the distribution of the relative hole sizes, $d/\Lambda$, for the four inner and nine rings in the cladding. The four inner rings show a narrow size distribution with an average relative hole size of 0.659 and a standard deviation of 0.024, whereas for the nine full rings, the distribution is broadened toward smaller holes, yielding an average of 0.627 and a standard deviation of 0.070. In the following, we will use the size distribution of the inner four rings to characterize the fiber geometry, since it was theoretically found, that the cut-off does not change when the number of rings is increased from four to nine.\cite{kuhlmey2002} In Table~\ref{tab1}, the pitch and relative hole size are listed for all the seven fibers reported on here.

\begin{table}[t!]
\begin{center}
\begin{tabular}{|c|c|c|c|c|c|c|c|}
\hline
fiber \# & 1 & 2 & 3 & 4 & 5 & 6 & 7\\
\hline
$\Lambda/{\rm \mu m}$ & 1.41 & 1.20 & 1.52 & 1.10 & 1.21 & 1.20 & 1.20\\
$d/\Lambda$& 0.66 & 0.64 & 0.68 & 0.63 & 0.59 & 0.57 & 0.54\\\hline
\end{tabular}
\end{center}
\caption{Pitches, $\Lambda$, and relative air-hole diameters, $d/\Lambda$, of the seven investigated fibers. Typically, the uncertainty on the measured $d/\Lambda$ is $\pm$ 0.02. }
\label{tab1}
\end{table}

For the experimental determination of the cut-off in the NL-PCFs, we have recorded the transmitted intensity in the fiber as a function of wavelength. The transmission shows a significant increase below the cut-off wavelength since more modes are available to carry power. This measurement principle is also used in standard fiber technology (see {\it e.g.} Ref.~\onlinecite{ghatak1998}), however due to the small cores and high index steps in the NL-PCFs special care must be taken to launch and collect light with sufficiently high NA to measure the contributions from all modes. In the set-up used here, white light from a Tungsten Halogen lamp is coupled into the fiber using a microscope objective (MO) with an NA of 0.9. At the output, the emerging light is collimated using an identical MO and focused into a multi-mode fiber with an MO having an NA of 0.16. The multi-mode fiber is connected to an optical spectrum analyzer. The chromatic aberrations of the lens pair at the output as well as the white light source itself give rise to a strongly wavelength dependent power spectrum. In order to obtain a reference spectrum, the measurements have been performed starting with 3.0 meters of fiber, which is then cut back to 0.5 meters keeping the output coupling constant.

\begin{figure}[b!]
\begin{center}
\epsfig{file=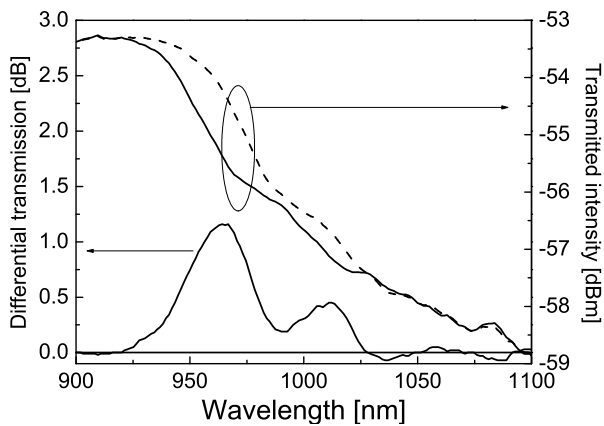, width=0.45\textwidth,clip}
\end{center}
\caption{Right axis: Transmission spectra of 0.5 meter (dashed line) and 3.0 meter (solid line) of fiber \#1. Left axis: Difference of the transmission spectra.}
\label{fig2}
\end{figure}

In Fig.~\ref{fig2} the transmission spectra are shown for the two lengths of fiber \#1, as well as the difference of the spectra. Since the higher order modes are very lossy close to cut-off, a significant difference is expected between the transmission on the long length and the short length of fiber,\cite{argyros2001} which is clearly seen in the spectral range between 925 nm and 1025 nm. As an alternative to using the cutback technique, we also tried keeping the length constant, recording the transmission spectra for different bending radii of the fiber, as prescribed in several standard fiber measurement procedures.\cite{CEI/IEC} However, no measurable difference between the spectra were observed for radii down to less than 5 mm. Most likely this is due to the high NA of the higher-order modes that makes them much less sensitive to bending compared to higher-order modes in standard fibers.

\begin{figure}[t!]
\begin{center}
\epsfig{file=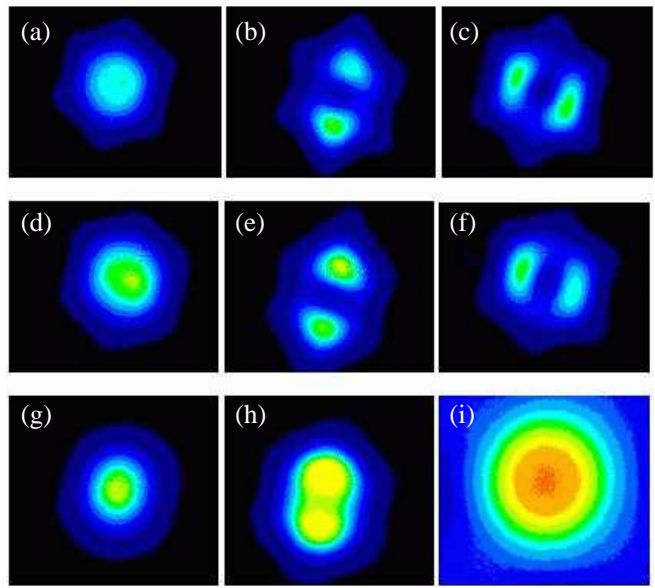, width=0.48\textwidth,clip}
\end{center}
\caption{Near field images recorded at the output of 2.0 meters of fiber \#1. The different panels show (a) the fundamental mode at 635 nm, (b,c) higher-order modes at 635 nm, panel (d) fundamental mode at 780 nm, (e,f) higher-order modes at 780 nm, (g) fundamental mode at 987 nm, (h) the only observed higher-order mode at 987 nm, and (i) fundamental mode at 1550 nm.}
\label{fig3}
\end{figure}

The double peak structure of the subtracted spectra was observed in all the fibers investigated here. In order to clarify the nature of the double peak we investigated the spatial structure of the modes propagating in 2 meters of fiber at 635 nm, 780 nm, and 987 nm. These results are summarized in Fig.~\ref{fig3}. Both at 635 nm and 780 nm it is possible to excite a fundamental mode with a maximum in the middle of the core [panels (a) and (d)], as well as two spatially orthogonal double-peaked intensity profiles having a minimum in the middle of the core [panels (b), (c), (e), and (f)]. However, at 987 nm only one of the double-peaked intensity profiles could be excited [panel (h)], which was observed independent of the orientation of the polarization of the launched light.

The source at 987 nm is located in the minimum between the two peaks in Fig.~\ref{fig2} and hence each of the peaks are attributed to higher-order modes [with the profiles in {\it e.g.} panels (e) and (f)] having different cut-off wavelengths, in this case at 1012 nm and at 970 nm. In the ideal NL-PCF structure the first four higher order modes are in fact not four-fold degenerate,\cite{steel2001} as they are in the weakly guiding approximation for circular step index fibers.\cite{snyder} In the presence of weak asymmetry, plane wave simulations show that the mode-indices of the four high-order modes group two by two. In the first group the modes have a spatial structure similar to panels (b,e,h) and the two modes in the other group have a structure corresponding to panels (c,f). Within each group we find that the two modes are orthogonally polarized, in agreement with the experimental findings. The presence of a small asymmetry in Fiber \#1, is supported by measurements of the group birefringence of the fundamental mode at 1050 nm, yielding a value of $\Delta n = 1.3\times 10^{-4}$ (note that $\Delta n=0$ for the ideal structure\cite{steel2001}). With the small pitches in the fibers investigated here, only a very small variation of the pitch or hole size around the core may give rise to a birefringence on this order of magnitude.\cite{nielsen2001}

\begin{figure}[t!]
\begin{center}
\epsfig{file=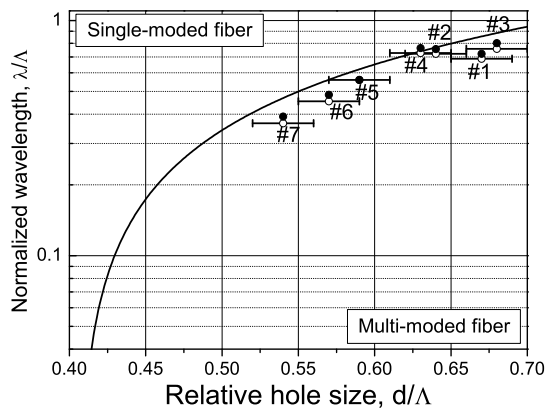, width=0.45\textwidth,clip}
\end{center}
\caption{The plotted points show the relative cut-off wavelength, $\lambda_{\rm cut-off}/\Lambda$, as a function of the relative hole-size of the fibers investigated here. The solid (open) circles indicate the short (long) wavelength cut-off. The solid line shows Eq.~(\ref{cut-off}).}
\label{fig4}
\end{figure}

All the cut-offs of the fibers measured here are plotted in Fig.~\ref{fig4} as a function of relative hole size in units of the normalized wavelength, $\lambda/\Lambda$. The solid line shows the phenomenological relation

\begin{equation}\label{cut-off}
\lambda_{\rm cut-off}/\Lambda\simeq \alpha (d/\Lambda-0.406)^{\gamma}
\end{equation}
where $\alpha\simeq 2.80\pm 0.12 $ and $\gamma\simeq 0.89\pm 0.02$. This relation was found to account well for numerical cut-off simulations\cite{kuhlmey2002} and recently we have shown that it can also be derived from $V$-parameter considerations.\cite{mortensen2003c} These predictions are in very good agreement with observed cut-off dependence, although the experimental data generally yield a slightly lower cut-off wavelength. It should be noted that the experimental criteria for cut-off is different than the criteria used in the simulations. The measurements described here defines the cut-off as the wavelength at which the attenuation of the higher order modes is on the order of 1 dB/m, whereas the simulations use the peak of the second derivative of the attenuation to define the cut-off. 

In conclusion, we have demonstrated a measurement procedure for determining the modal cut-off in NL-PCF, which takes into account the high NA of the modes as well as the weak sensitivity to bends. The method is validated by inspecting the near field of the modes that may be excited in the fiber above and below cut-off. The cut-off spectra show a double-peak structure, which is attributed to a splitting of the higher order modes induced by asymmetry in the structure, in agreement with a high birefringence observed for the fundamental mode. Seven different NL-PCF designs with varying pitches and relative hole sizes have been investigated, and very good agreement is found with recent theoretical work on the modal cut-off.

\vspace{5mm}
N.~A. Mortensen thanks B.~T. Kuhlmey for discussions important to this work and K.~P. Hansen and T.~P. Hansen acknowledge financial support by the Danish Academy of Technical Sciences. J.~R. Folkenberg's e-mail address is jrf@crystal-fibre.com.

\end{document}